\documentclass[pre,aps,floatfix]{revtex4}

\usepackage{amsmath,amssymb}
\usepackage{graphicx}
\usepackage{psfrag}
\usepackage{graphicx}

\def\beq{\begin{equation}}
\def\eeq{\end{equation}}
\def\bea{\begin{eqnarray}}
\def\eea{\end{eqnarray}}

\begin{document}

\title{Instabilities and diffusion in a hydrodynamic model of a fluid membrane coupled to a thin active fluid layer}
\author{Niladri Sarkar and Abhik Basu} \affiliation{{Theoretical Condensed Matter
Physics Division, Saha Institute of Nuclear Physics, 1/AF,
Bidhannagar, Calcutta 700 064, India} }

\date{\today}

\begin{abstract}
We construct a coarse-grained {\em effective} two-dimensional ($2d$)
hydrodynamic theory as a theoretical model for  a coupled system of
a fluid membrane and a thin layer of a {\em polar active fluid}  in
its ordered state that is anchored to the membrane. We show that
such a system is prone to generic instabilities through the
interplay of nonequilibrium drive, polar order and membrane
fluctuation. We use our model equations to calculate diffusion
coefficients of an inclusion in the membrane and show that their
values depend strongly on the system size, in contrast to their
equilibrium values. Our work extends the work of S. Sankararaman and
S. Ramaswamy [{\em Phys. Rev. Lett.}, {\bf 102}, 118107 (2009)] to a
coupled system of a fluid membrane and an ordered active fluid
layer. Our model is broadly inspired by and should be useful as a
starting point for theoretical descriptions of the coupled dynamics
of a cell membrane and a cortical actin layer anchored to it.
%
\end{abstract}

\maketitle

\section{Introduction}



Theoretical studies of equilibrium properties (both static and
time-dependent) of fluid membranes are extensive and
well-established by now~\cite{seifert-rev,helfrich,sriram1}.
Mechanical properties of membranes are essentially controlled by a
few parameters, such as the membrane tension, bending modulus and
spontaneous curvature~\cite{seifert-rev}. These parameters
completely characterise  membrane fluctuation spectra at thermal
equilibrium. By contrast, theoretical studies of the dynamical
properties of driven fluid membranes are relatively at early stages
of development. In this article, we consider the coupled dynamics of
a single-component fluid membrane and an active fluid layer anchored
to it, within a coarse-grained hydrodynamic approach. Apart from its
direct theoretical interests, there are biological implications as
well. Studies of biomembranes are important in cell biology context,
since all cells are covered by cell membranes which typically have
complicated internal structures \cite{alberts}. At a coarse-grained
level, however, cell membranes are usually described as a bilayer
fluid (with complicated microscopic internal structures, see, e.g.,
\cite{singer,alberts}). In eukaryotic cells, there is a thin layer
of cross-linked actin filaments anchored to the membrane and
associated with the cell cytoskeleton in typical eukaryotic cells.
Knowledge of the properties of fluid membranes in equilibrium are
not sufficient to characterise real biological membranes, such as
plasma membranes of any cell, which are inherently nonequilibrium
{\em active} systems \cite{alberts,sriram-madan,saxton}. For
instance, cell membranes are constantly maintained out of
equilibrium due to nonequilibrium processes of various types. It has
now been realized that such nonequilibrium behaviour may underlie
aspects of biomembrane dynamics \cite{levin,tuvia}, previously
attributed purely to
equilibrium thermal fluctuations. 
These nonequilibrium aspects affect experimental measurements of
physical quantities characterising cell membrane dynamics, e.g., one
finds anomalous diffusion \cite{saxton,eddin} (however, these are
primarily due to coralling and inhibiting effects, different from
active fluid effects discussed here).
It is now generally accepted that such complex dynamical behaviour
have their origins in generic nonequilibrium drives. In a living
cell, there may be several different sources for such nonequilibrium
behaviour - it could be due to active proteins inside the membrane
that are often Adenosine Triphosphate (ATP) consuming enzymes, or
the membrane could be driven out of equilibrium by the
nonequilibrium fluctuations of an active cortical actin layer
anchored to the membrane. The former case has been studied both
experimentally and theoretically in details, see, e.g.,
Ref.~\cite{manne}. The latter case is discussed in
Refs.~\cite{nir2,falcke}. In particular, Ref.~\cite{nir2} introduced
a model that couples a membrane with actin polymerisation and
contractile forces due to molecular motors, e.g., myosin. For a
configuration of actin filaments which are typically normally
directed to the membrane, Ref.~\cite{nir2} has shown how motor
activities lead to generic transverse membrane waves.
Ref.~\cite{falcke} discusses the influence of actin filament
elasticity and retrograde flow on the force-velocity relations of
motile cells. A related problem is the diffusion of small inclusions
embedded in membranes, which is challenging both theoretically and
experimentally. This has been extensively studied. For instance,
Ref.~\cite{nir1} theoretically shows how curvature fluctuations
(both thermal and non-thermal) in a membrane leads to enhancement of
the effective thickness, that in turn causes a reduction in the
measured diffusion coefficient. In a related work, Ref.~\cite{naji}
discusses how membrane fluctuations, both static or quenched and
annealed or dynamic, affect the diffusion coefficient. By using a
mixture of analytical and numerical methods they find that for a
membrane with quenched fluctuations, the diffusion coefficient is
substantially reduced by the quenched roughness, and is nicely
described by an area-scaling law proposed there. In contrast, for
the annealed case with small bending rigidity, the reduction in the
the diffusion coefficient is less than in the quenched case. These
studies on diffusion coefficients, however, do not include any
nonequilibrium effects that arise due to the {\em active stress}
(see below) in the active fluid layer anchored to the membrane.
Elucidation of the generic effects of the active stress present in
the ordered active fluid layer on the statistical properties of the
attached fluid membrane is a major purpose of the present study.

Although a cell cortex is typically not in an ordered state for the
polarisation degrees of freedom (the actin filaments), we consider
an ordered state and study fluctuations about it. In
particular, we assume the ordering direction to be parallel to the
membrane. The other possible ordered state, {\em viz} where the
actin filaments are predominantly perpendicular to the membrane is
important as well (see, e.g., Ref.~\cite{nir2} for a study with
perpendicular configuration of the filaments). Since a prime
motivation of this work is to theoretically study in a general set
up the effects of the nonequilibrium dynamics and fluctuations of an
ordered layer of polar active fluid on a fluid membrane that is in
contact with the active fluid layer, we have chosen to work with the
parallel orientation as an illustrative example. Nevertheless, our
scheme of calculations may be easily extended to filaments with
perpendicular ordering. We briefly discuss this later. Our work
here will have broad relevance and act as a starting point for more
detailed study of the dynamics of a single-component fluid membrane
in contact with a thin layer of cortical actin filaments. Recently,
artificial actin layers attached to a fluid membrane have been
constructed and experiments performed on them \cite{bass,tsai}.
Experiments include investigation of the structure of the actin
layer and measurement of diffusion of inclusions in the membrane.
The latter experiments do find rising lateral diffusion coefficients
for larger systems \cite{bass}. While preparation of and
experimental studies on artificial actin layers of a particular
macroscopic configuration (i.e., a particular ordered state)
anchored to a membrane is definitely technically a difficult task,
some of our results described here may in principle be tested by
performing controlled experiments on such systems.

In this article, a thin cortical layer of actin filaments is
considered at length scales much larger than the filament lengths
and layer thickness with polar ordering, for which a generic
coarse-grained continuum two-dimensional ($2d$) description would be
appropriate. Further, we consider time scales larger than the
unbinding time scale of the cross-linking proteins of the actin
filaments, so that the actin network behaves like a fluid. In order
to model the nonequilibrium dynamics of cortical actin layer, we use
the {\em active fluid} description, proposed and elucidated in
Refs.~\cite{sriram2,kruse}. Such approaches have been successfully
used in a variety of driven systems, see, e.g.,
Refs.~\cite{sriram-rev,gautam-rev,joanny-rev,sriram-rmp} for
extensive discussions about the subject.  There have been a number
of studies on the properties of active fluid in confining thin
geometries before. For example, Refs.~\cite{salbreux,mayer} proposed
a mechanism for development of contractile rings in cell cortex.
Ref.~\cite{mayer} discussed large scale flows of the acto-myosin
cortex in {\em Caenorhabditis elegans} zygotes. Further,
Ref.~\cite{frank} provides a one-dimensional ($1d$) model for
pattern formation in active fluids (see also Ref.~\cite{alex} for a
$1d$ model of patterns in a cell cortex). In addition,
Ref.~\cite{dober} points out the existence of universal dynamic
patterns by lateral membrane waves in motile cells. Our results here
are complementary to those mentioned above. In this article, we
systematically derive a set of {\em effective} $2d$ coarse-grained
hydrodynamic equations of motion for the membrane height field,
local orientational order parameter, concentration and the velocity
fields, which we use to obtain results characterising fluctuations
and correlations in the system. In particular, we show that there
are generic moving or static instabilities (i.e., with or without
propagating modes). 
We use a stochastically driven versions of these to illustrate the
effects of the active (nonequilibrium) dynamics on the lateral and
rotational diffusion coefficients of an inclusion in the membrane.
We find that under certain situations there could be non-trivial
system size dependences of the diffusion coefficients in comparison
with their equilibrium expressions. The rest of this article is
organised as follows: In Sec.~\ref{model} we describe the basic
model considered here and set up the equations of motion. We then
show in Secs.~\ref{instabilI} and \ref{instabilII} that these
equations, upon linearisation show emergence of linear instabilities
when the strength of the nonequlibrium drive, denoted by $\Delta\mu$
exceeds a critical value. Finally, in Sec.~\ref{diff} we calculate
lateral and rotational diffusion coefficients of an inclusion in the
membrane; we show that for a membrane-active fluid combine lateral
and rotational diffusion coefficients depend on the system size in
ways that are very different from the known results for diffusion in
systems at equilibrium. We summarise and discuss our results in the
context of existing experimental results in Sec.~\ref{conclu}.

\section{Model system and equations of motion}
\label{model}

For the sake of simplicity, we analyse here the effect of  activity
on the fluctuation spectrum of a planar (flat on an average)
membrane only. Our calculational framework as we formulate below is
general. However, we use it to study, within a linearised approach,
typically the properties of perturbations about an ordered, uniform
reference
state of the active fluid.
 The construction of our model is generally inspired by the structure
of a real eukaryotic cell. There is a thin layer of cortical actin
or {\em cell cortex} anchored to the cell membrane in an eukaryotic
cell. This layer is rich in actin filaments which may be locally
preferentially orientated either parallel or perpendicular to the
membrane.  
The bulk of the cell is the cytoplasm consisting of cytoskeleton and
other cellular organelles.
We model the cortical actin layer by a thin film of active fluid of
viscosity $\eta$, thin in the $z$-direction and spread along the
$xy$ plane with active polar particles (since the actin filaments
are polar) suspended in it. In our discussions below we only
consider a macroscopically ordered state of the active particles
(actin filaments) with in-plane ordering. We do not discuss the
other possible case of perpendicular orientations in details here.
The active fluid is assumed to be anchored to one side of a fluid
membrane characterised by a bending stiffness. We consider two
distinct versions of our system: (i) Model I: The fluid
membrane-active fluid combine is embedded inside a bulk isotropic
fluid on both sides. The bulk fluid viscosity $\eta'$ is assumed to
be much smaller than the active fluid viscosity (for simplicity; see
below), and (ii) Model II: The system rests on a solid substrate
below. The two cases are physically different: First of all, the
presence of the solid substrate introduces friction and, as a
result, the momentum (same as the hydrodynamic velocity ${\bf v}$
for an incompressible system) is no longer a conserved field. In
contrast, when there is a bulk fluid surrounding the system there is
no friction at the interface and hence $\bf v$ remains a conserved
variable. This leads to {\em long-ranged hydrodynamic interactions}
in such a system. Secondly, the presence of the solid substrate
breaks the full three-dimensional ($3d$) rotational invariance of
the problem, where as, when there is an embedding bulk fluid, the
system is invariant under the full $3d$ rotational invariance (see
below). This has important ramifications on the possible structure
of the membrane free energy, as we will discuss below. Despite the
cell biological inspiration for our work, the connections between
our model and the structure of an eukaryotic cell is not very
strict, and thus our model and results cannot be directly applied to
a biologically relevant system: First of all, the cortical actin
layer does not exist in an ordered state with in-plane orientation;
at the scale of the cell it is generally isotropic. Secondly, in a
real cell, there is not any sharp dividing surface separating the
cortex from the bulk. Although our idealised model is not accurate
for a real cell, nevertheless it allow us to obtain interesting and
non-trivial result in a simple set up, highlighting how the {\em
environment} (in the form of a solid substrate below or a passive,
isotropic embedding fluid around the system) may drastically affect
macroscopic physical properties.  Our work may be considered as a
starting point of more realistic calculations.
Our treatment generalises Ref.~\cite{sumithra} to the case of a
coupled system of a fluid membrane and an active fluid film. (See
also Ref.~\cite{friedrich} for discussions on the effects of a solid
substrate on the dynamics of thin liquid film covered by a
membrane.) We now discuss the constructions of Model I and Model II
in some details below.

\subsection{Model I: System covered by isotropic fluid on both
sides} \label{modelI}

Consider first our Model I, where  the fluid membrane-active fluid
layer combine is embedded by an isotropic fluid on both sides. In
order to keep the ensuing algebra tractable and to obtain closed
equations and non-trivial results without having to introduce too
many details we make a number of simplifying assumptions which we
now discuss. First of all, we assume $\eta \gg \eta'$. To set up the
context for our work, we refer to the well-known results in
Ref.~\cite{saffman}, where the translational diffusivity of an
inclusion is calculated in a thin $2d$ inflexible flat layer of an
isotropic viscous fluid of viscosity $\eta$ confined between bulk
isotropic viscous fluids of viscosity $\eta_1$ ($\eta_1/\eta$
finite) on both sides. We here consider a thin active fluid layer
containing orientational degrees of freedom in their ordered state
and covered by a fluid membrane on one side of finite stiffness.
However, we consider only the special limit of $\eta \gg \eta'$.
Thus our work here can be thought of as generalisation of and
complementary to Ref.~\cite{saffman}.

\begin{figure}[htb]
\includegraphics[height=6cm]{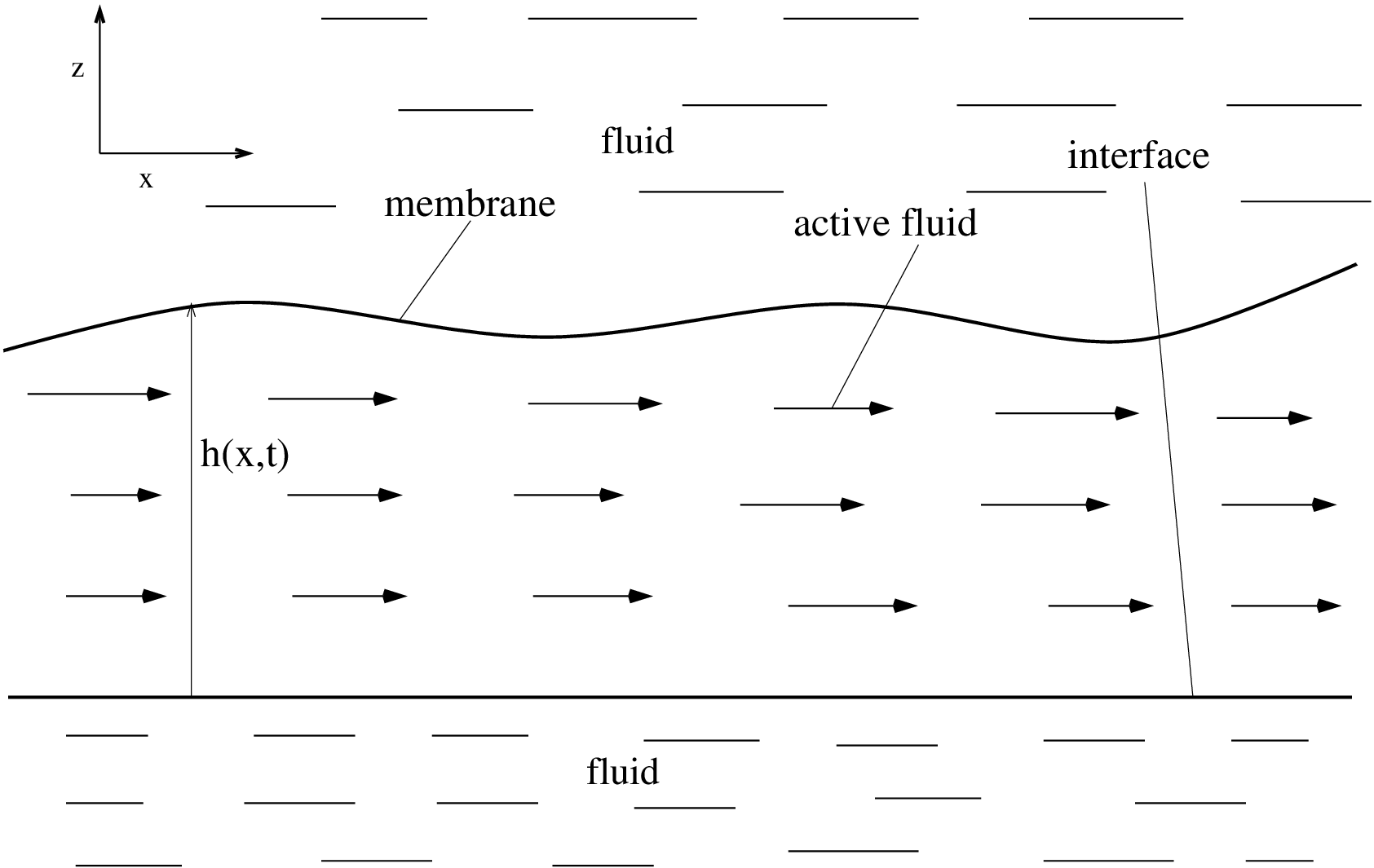} \hfill
\caption{A schematic diagram of our model system showing the
membrane and the active fluid layer.  The arrows indicate the
direction of macroscopic orientation (here along the $x$-axis).}
\label{scheme1}
\end{figure}

In order to specify the problem completely, we impose the following
boundary conditions: When the fluid membrane-active fluid layer is
covered by an isotropic fluid on both sides, the boundary conditions
are as follows: (i)  At the  interfaces ($z=h_1\; {\rm and}\; h_2$)
we impose $\bf p$ to be parallel to the local tangent plane on the
surfaces, i.e., ${\bf p}\cdot \hat{N} =0$ at $z=h_1$ and $h_2$, as
the current of active particles is along ${\bf p}$ and the particles
cannot leave the film, where $\hat N$ is the local normal at $h_1$
and $h_2$, and (ii) continuity of the shear stress at $z=h_1\; {\rm
and}\; h_2 $. The free energy functional $\mathcal F_p$ of the
system is a functional of $h_1,\,h_2$ and $\bf p$. The form of
$\mathcal F_p$ may be inferred from symmetry considerations. It must
generally be invariant under an arbitrary tilt (equivalently a
rotation) $ h_{1,2}\rightarrow h_{1,2} + {\bf a\cdot x}$ of the free
surfaces, where $\bf a$ is an arbitrary $3d$ vector and $\bf x$ is a
$3d$ radius vector. This ensures that the most leading order (in
gradients) coupling bilinear in $\bf p$ and $h_1$ or $h_2$ could be
of the form ${\boldsymbol\nabla}\cdot {\bf p} \nabla^2 h_{1,2}$.
This is a polar term, since it has no $\bf p\rightarrow -p$
symmetry. Further, it violates $h_{1,2}\rightarrow -h_{1,2}$
symmetry as well, which is admissible since the actin filaments are
anchored only on one side of the membrane. For the polar order
parameter, we use the Frank free energy \cite{prost} in the limit of
equal Frank's constants, denoted by $D$ here. Assuming that the
surface tension of the membrane is negligible, the general form of
the free energy functional of the combined system of membrane (top),
free surface (bottom) and active fluid is given by (in the Monge
gauge \cite{weinberg})
 \bea \mathcal{F}_L[h_1,h_2,{\bf p}] &=& {1 \over 2}\int d^2r
 \left[\sigma ({\boldsymbol\nabla}_\perp h_2)^2 +
 \kappa(\nabla_\bot^2 h_1)^2\right] \nonumber \\
&+& {1\over 2} \int d^2 r \int_{h_2}^{h_1} dz\left[\hat{C}
({\boldsymbol\nabla}\cdot{\bf p}) (\nabla^2 h_1) \delta (z-h_1) +
\hat C ({\boldsymbol\nabla}\cdot {\bf p})\nabla^2 h_2\delta (z-h_2)+
D (\nabla_i p_j)^2\right], \label{freeIA}\eea
 where $\kappa$ is the bending rigidity of
the membrane, $\sigma$ the interfacial tension of the surface at
$h_2$, $\hat{C}$ is a coupling constant that couples the
orientational field ${\bf p}$ with the fluctuations of the height
fields. In principle, one needs to solve for the dynamics of $h_1$
and $h_2$ separately for a full dynamical description of the
problem. Operators ${\boldmath
\nabla}_\perp=(\frac{\partial}{\partial x},\frac{\partial}{\partial
y})$ and ${\boldmath \nabla}=(\frac{\partial}{\partial
x},\frac{\partial}{\partial y},\frac{\partial}{\partial z})$ are the
$2d$ and $3d$ Laplacians, respectively. At this stage, in order to
simplify the ensuing algebraic manipulations, but still be able to
display the dramatic effects of the hydrodynamic interactions, we
consider a special simplifying (but admittedly artificial) limit of
the problem with $\sigma$ being large enough (formally a diverging
$\sigma$), such that fluctuations of $h_2$ are suppressed and may
henceforth be ignored. Thus we are required to solve for the
dynamics of $h_1$ only (along with $\bf p$ and $c$). It must be
mentioned that the limit of large $\sigma$ is primarily of
theoretical interests and is not expected to be observed in any real
(biological) system. Despite these limitations of our assumption of
large $\sigma$, we obtain interesting results which are expected to
be present qualitatively even with finite $\sigma$. In order to set
notations simpler, we set $h_1=h$ and $h_2=0$ in our subsequent
analysis below. Thus $h_2$ drops out of the dynamics; see
Fig.~(\ref{scheme1}) for a schematic picture of our Model I system.
Consequently, the free energy (\ref{freeIA}) reduces to \bea
\mathcal{F}_L[h,{\bf p}] &=& {1 \over 2}\int d^2r \left[ \kappa(\nabla_\bot^2 h)^2\right] \nonumber \\
&+& {1\over 2} \int d^2 r \int_{0}^{h} dz\left[\hat{C}
({\boldsymbol\nabla}\cdot{\bf p}) (\nabla^2 h) \delta (z-h) +D
(\nabla_i p_j)^2\right].\label{freeI}
\end{eqnarray}
It is invariant under $h\rightarrow h + {\bf a\cdot x}$.
 Now imposing the kinematic boundary conditions
\cite{kine} on $v_z$ at $h$ for an impermeable membrane we write:
 \begin{equation}
{\partial h \over
\partial t} + {\bf v_\perp}\cdot{\boldsymbol\nabla}_\perp h=v_z,\label{heq}
\end{equation}
 which connects the height fields $h$  with the
hydrodynamic velocity field ${\bf v}=(v_x,v_y,v_z)$ at the location
of the membrane ($z=h$). Here, ${\bf v}_\perp=(v_x,v_y)$. In the
expression (\ref{freeI}) above, when $D-\hat C^2/\kappa>0$, the
equilibrium phases are spatially uniform, else modulated phases are
possible. In this article we consider $D-\hat C^2/\kappa > 0$,
corresponding to uniform equilibrium phase only \cite{lub1}. Thus
finally the relevant slow modes in this {\em effective} $2d$ problem
are (i) a concentration of active polar particles $c({\bf r},t)$,
(ii) a local orientation field (describing the local orientation in
the cortical actin) given by ${\bf p}({\bf r},t)$, and (iii) a local
height fields $h({\bf r},t)$ to describe the nearly flat membranes
(in the Monge gauge \cite{weinberg}), where ${\bf r}=(x,y)$ is the
in-plane coordinate and $t$ is the time. We take the $xy$-plane as
the easy plane for polarisation fluctuation. Further, we consider
macroscopic orientational order given by a reference state ${\bf
p}=(1,0,0)$ i.e. we have macroscopic ordering along the $x$ axis and
impose fixed length constraint on $\bf p$: $p^2=1$. For an active
system, the polarity implies a current $v_0c{\bf p}$ with respect to
the fluid, where $v_0$ is a characteristic drift velocity.

 The relevant intrinsic
stress field of the active particles is given by
 \cite{joanny-rev,sriram-rev,sumithra}
\bea
\sigma^a_{ij}=\Delta\mu
c({\bf r}) p_i ({\bf r}) p_j ({\bf r}),\label{stress}
\eea
 which is of nonequilibrium origin. Stress $\sigma^a_{ij}$ is said to be contractile or
 extensile for the constant $\Delta\mu <0$ or $\Delta\mu >0$. The generalised Stokes Equation for $\bf v$, is obtained by using the
force balance condition after neglecting inertia as appropriate for
small masses in typical biological systems. Velocity components $\bf
v_\perp$ are to be solved from the full Stokes equation
 \bea \eta\nabla_\bot^2 {\bf v}_\bot +
\eta\partial_z^2 {\bf v}_\bot - {\boldsymbol\nabla}_\bot \Pi -
\nabla_j\sigma^a_{\bot j} =0,\label{stokesxyI}
 \eea
where $\sigma^a_{\perp j}=(\sigma^a_{xj},\sigma^a_{yj})$.
At the membrane ($z=h$) we have ${\bf p}\cdot \hat{N} =0$ where
$\hat N$ is the local normal at $h(x,y)$. By using the Monge gauge
for the membrane we have $\hat{N}=(-{\boldsymbol\nabla}_\bot
h,1)/\sqrt{1+(\nabla_\bot h)^2}$ to be the outward normal to the
membrane surface. This gives us
\begin{equation}
p_z\simeq \frac{\partial_x (h)}{h}z.\label{interpz}
\end{equation}
This then yields $\partial_z p_z\simeq {1 \over {h}}\partial_x h$.
In order to solve for ${\bf v}$ from the above equations we must
first find out ${\boldsymbol\nabla}\cdot \sigma^a$. Within our
linearised treatment, assume ${\bf p_\perp}=\hat x+ \theta \hat y$
with small $\theta$, where ${\bf p_\perp}= (p_x,p_y)\sim (1,\theta)$
to the lowest order. With $|\theta|\ll 1$ considering small
fluctuation, we may obtain different components of $\nabla_j
\sigma^a_{ij}$.
To eliminate $\Pi$ we solve the Stokes equation for $v_z$ in the
lubrication approximation, i.e., we write $\partial_z
\Pi=-\partial_i\sigma^a_{iz}=-\Delta\mu {c_0 \over 2}\partial_x^2
h$. This yields
 \bea \Pi(x,y,z,t) = P_0 +
\Delta\mu {c_0 \over 2} (h-z)\partial_x^2 h -f(h), \label{pressureI}
\eea
 where $f(h)=-{\delta\mathcal{F_L} \over \delta h}= -
\kappa\nabla_\bot^4 h + \hat{C}
\nabla_\bot^2{\boldsymbol\nabla}_\bot \cdot {\bf p}_\bot$ is the
elastic force of the membrane. Equation~(\ref{pressureI}) tells us
that the balance of the normal component of stress at the membrane
gives the condition of balancing the fluid stress by the elastic
force of the membrane (\ref{freeI}). Thus the active
contribution to the pressure in (\ref{pressureI}) comes with the
same signature of $\Delta\mu$ as in the active stress expression
(\ref{stress}).

The solutions of equation (\ref{stokesxyI}) are facilitated greatly
in terms of the in-plane Fourier transformation: For a system as
above, the general form for the hydrodynamic kernel is $1/(\eta q^2
+ \eta' q/h_0)$ where $h_0$ is the average thickness of the system,
where $\eta$ and $\eta'$ are the viscosities of the active fluid and
the surrounding passive isotropic fluid, respectively. Thus for $q>
\eta'/(\eta h_0)$, the hydrodynamic kernel is identical to that of a
free standing system. If (for the sake of simplicity) we now assume
that $\eta \gg \eta'$, then in a broad window of wavevector, the
hydrodynamic kernel may be approximated by $1/\eta q^2$. Further in
that limit, the continuity of the shear stress at $z=h$ and $z=0$
actually implies vanishing of the shear stress at $z=h$ and $z=0$.
Hence, in order to solve for $v_x$ and $v_y$, we now need to solve
Eq.~(\ref{stokesxyI}) subject to boundary conditions of zero shear
stress at $z=h$ and $z=0$. To the leading order in smallness this
translates into
\begin{equation}
\eta \frac{\partial v_x}{\partial z} + \Delta\mu
p_z=0,\;\;\eta\frac{\partial v_y}{\partial z}=0,
\end{equation}
at $z=h$ and
\begin{equation}
\eta \frac{\partial {\bf v}_\perp}{\partial z}=0
\end{equation}
at $z=0$ (since $p_z=0$ strictly at $z=0$).  Thus, for this
wavevector range, effectively we have a free standing active fluid,
covered on one side by a fluid membrane and a free surface on the
other. In what follows below we stay in this limit only
\cite{foot1}.  Thus
 \bea
 v_y= -i{q_y \over {\eta q^2}}\left[-{3c_0\Delta\mu h_0 \over
4}q_x^2 h + \kappa q^4 h -i\hat{C} q_yq^2\theta \right] -
i{c_0\Delta\mu \over \eta q^2}q_x\theta +f^L_y, \label{eqy}\eea
 \bea v_x &=& -{1 \over \eta q^2}\left[-i{3\Delta\mu h_0c_0  \over
4}q_x^3h + i\kappa q_xq^4h + \hat{C} h_0q^2q_xq_y\theta + i(\Delta\mu
q_x c +
 c_0\Delta\mu q_y\theta + {2c_0\Delta\mu \over h_0}q_xh) \right] +
f^L_x,\label{eqx}
 \eea
 for $qh_0 \ll 1$ (we work in this limit which enables us to continue using the
 Lubrication approximation).
 Here we have performed an in-plane Fourier transform with $\bf q$
 as the Fourier conjugate of $\bf r$. Functions $f^L_i,i=x,y$
 are zero-mean, Gaussian noises with variances $\langle f_i^L({\bf
 q},\omega)f_j^L(-{\bf q},-\omega)\rangle= \frac{2K_BT}{\eta q^2}\delta_{ij}$.
 We have considered thermal noises for simplicity, although in real
 biological situations there are non-equilibrium noises as well.

From the incompressibility of the fluid we obtain $v_z=-\int
{\boldsymbol\nabla}_\bot\cdot {\bf v}_\bot dz$ together with the
condition $v_z=0$ at $z=0$. Neglecting terms with higher derivatives
and linearising about $\langle h\rangle=h_0$ and $\langle
c\rangle=c_0$ and denoting $h$ and $c$ as the height and
concentration fluctuations from their respective averages we obtain
by using Eqs.~(\ref{eqy}) and (\ref{eqx})
 \bea
\frac{\partial h}{\partial t}=v_z(z=h) &=& {1 \over \eta q^2}[
{3\Delta\mu c_0 h_0^2 \over 4}q^2q_x^2h + i\hat{C}h_0q^4q_y\theta -
\kappa h_0q^6h -(\Delta\mu h_0 q_x^2 c \nonumber \\ &+&
2c_0h_0\Delta\mu q_xq_y\theta + 2\Delta\mu c_0q_x^2h)] +\xi_h
\label{eqh1}
 \eea
 in the Fourier space for a free standing system, $\xi_h$ is a zero-mean Gaussian
 white noise related to $f^L_i$ and hence with a variance $\frac{2K_BTh_0^2}{\eta}$. 

Dynamics of the polar orientation field $\bf p$ differ from that of
the more usual nematic director in that the equation of motion of
$\bf p$ now must include terms \cite{aditi-polar} which violates the
$\bf p\rightarrow -p$ symmetry of the nematic director field. Apart
from the usual terms \cite{aditi-polar} we can have a
symmetry-permitted spontaneous splay term \cite{sumithra} in the
free energy functional like $\mathcal{F}_{sp}\equiv -\int d^3x
\tilde{C}{\boldsymbol\nabla}\cdot {\bf p}$, where $\tilde{C}$ is a
parameter which depends on the concentration of activity i.e.
$\tilde{C}=\tilde{C}(c_0) + \tilde{C}'(c_0)\delta c + ......\equiv C
+ C'\delta c$. Such terms may arise essentially due to the generic
structural differences between the {\em head} and {\em tail} of the
polar molecules. These will contribute $-\Gamma{\delta
\mathcal{F}_{L} \over \delta {\bf p}}= - \Gamma
C'{\boldsymbol\nabla} c$ to the equation of motion for $\bf p$,
$\Gamma$ being a kinetic coefficient. The $3d$ equation of motion
for $p_y=\theta$ becomes
 \bea
\partial_t\theta=-a_1 v_0\partial_x\theta - \xi\partial_y c  +
\hat{C}\nabla^2\partial_yh \delta(z-h) + (\lambda A_{yx} - \Omega_{yx}) + D\nabla^2\theta +\xi_\theta, \label{theta}
 \eea
  where $\xi=\Gamma C'$, $A_{ij}={1 \over
2}(\nabla_i v_j + \nabla_j v_i)$ is the strain rate tensor and
$\Omega_{ij}={1 \over 2}(\nabla_i v_j - \nabla_j v_i)$ is the
vorticity tensor.  The first term of Eq.~(\ref{theta}) represents
the advection of polar particles with the embedding fluid. The
coefficient $a_1$ is not necessarily unity, due to the lack of
Galilean invariance. The next two terms are coupling terms coupling
$\theta$ with the gradient of concentration of the active particles
and the fluctuation of the membrane surface respectively. Coupling
constant $\lambda$ couples flow (strain rate tensor) with the local
orientation. For stable flow-alignment $|\lambda| >1$ \cite{prost}.
The last term is the diffusion term representing the Frank free
energy \cite{frank} contribution to the dynamics in the limit of
equal Frank's constants. Noise $\xi_\theta$ is zero-mean, Gaussian
distributed with a variance $\frac{2K_BT}{\eta}$~\cite{foot-vari}.
After eliminating $v_x$ and $v_y$ by using Eqs~(\ref{eqx}) and
(\ref{eqy}), under $z$-averaging yields an equation for $\theta$,
given by equation (\ref{thetaeq}), as given in the Appendix (see
Sec.~\ref{appen1}) which
 depends only on $h,\,\theta$ and $c$. 
 When there is a solid substrate below, $\theta$ follows the same
 equation as (\ref{theta}), except the tilt-polarisation coupling
 term now reads $\hat C \partial_y h$.

In our system the drift velocity of the polar particles with respect
to the fluid is $v_0$ and hence the current due to active particles
is given by $v_0c{\bf p}$. Thus the continuity equation apart from
diffusion is given by $\partial_t c + {\boldsymbol\nabla}\cdot[({\bf
v}+v_0{\bf
p})c]=0$. Now 
 linearising about $c=c_0+c$
and using incompressibility we get \bea
\partial_t c = -iv_0c_0q_y\theta -iv_0q_xc +O(q_x^2,q_y^2). 
\label{conc} \eea 
\subsection{Model II: Membrane-active fluid combine rests on a solid
substrate}\label{modelII}

\begin{figure}[htb]
\includegraphics[height=6cm]{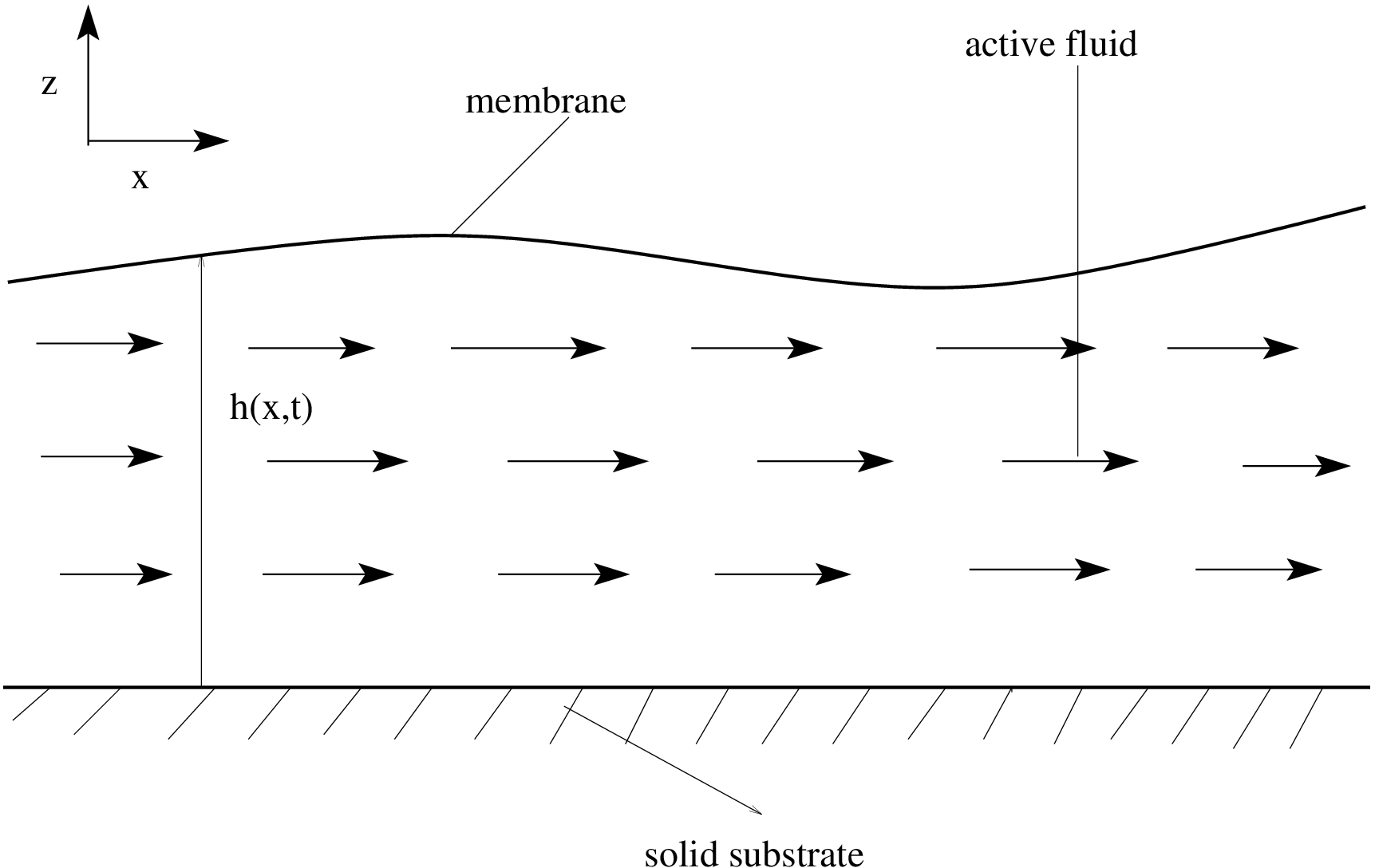}
\caption{A schematic diagram of our model system showing the
membrane and the active fluid layer.  The arrows indicate the
direction of macroscopic orientation (here along the $x$-axis).}
\label{scheme2}
\end{figure}
We now briefly discuss setting up the equations of motion when the
fluid membrane-active fluid layer combine rests on a fully flat
solid substrate; see Fig.~(\ref{scheme2}) for a pictorial
representation. Model II is important in the context in cell
locomotion on a rigid substrate and lamellipodium. A lamellipodium
is a cytoskeletal projection on the mobile edge of a cell. It is
effectively a quasi-$2d$ network of actin filaments and is
responsible for cell movement along a solid substrate (see, e.g.,
Ref.~\cite{alberts} for a general description of this). We
essentially follow the line of arguments and calculations of the
previous section, differing only in some details.  First of all, in
the present case we are concerned with the dynamics of the top fluid
membrane (at $z=h$) - active fluid layer combined, where as the
bottom solid surface (at $z=0$) is fixed. The latter is the
preferred frame of reference for this model (hereafter Model II).
This breaks the $3d$ rotational invariance discussed for Model I
above. This consideration then dictates the form of the free energy
$\mathcal F_S$ as \bea
 \mathcal{F}_S[h]={1 \over 2}\int d^2r   \kappa(\nabla_\bot^2 h)^2
+\frac{1}{2} \int d^2r \int_0^h dz\left[ \hat{C}
 {\bf p}\cdot {\boldsymbol\nabla} h\delta (z-h) + D(\nabla_i p_j)^2\right],\label{freeII}
\eea where, as before in (\ref{freeI}),  $\kappa$ is the bending
rigidity of the fluid membrane. Note that the bilinear coupling term
in $h$ and $\bf p$ is more relevant (in a renormalisation group
sense) than the corresponding terms in (\ref{freeI}). We continue to
denote the coupling constant by $\hat C$. Further we use the equal
Frank's constant limit, represented by $D$, for the Frank free
energy. The pressure $\Pi$ may be simply evaluated as above; we
obtain \bea \Pi= \Delta\mu {c_0 \over 2}(h-z)\partial_x^2 h +
\kappa\nabla_\bot^4 h  - \hat C{\boldsymbol\nabla}_\perp\cdot {\bf
p}_\perp, \label{pressureII}
 \eea
where we have used the boundary condition $\Pi=0$ at $z=0$ (setting
the zero of pressure) and balanced the normal component of stress at
$z=h$. Velocity components $v_x$ and $v_y$ may be calculated from
the Stokes Equation. For no-slip boundary conditions for $\bf
v_\perp$ at
 $z=0$, which we impose here,
 Stokes equation for $\bf v_\perp$ may be written in the lubrication
 approximation ($|{{\boldsymbol\nabla}_\bot \bf v_\perp}|<<|\partial_z \bf
 v_\perp|$)
 \bea
 \eta\partial_z^2 {\bf v}_\bot - {\boldsymbol\nabla}_\bot \Pi -
\nabla_j\sigma^a_{\bot j} =0.\label{stokesxyII}
 \eea
Now use (\ref{pressureII}) for $\Pi$ in
Eq.~(\ref{stokesxyII}) to calculate $v_x$ and $v_y$: Equation (\ref{stokesxyII})
may be twice integrated with respect to $z$ to obtain $v_x$ and
$v_y$. The ensuing constants of integration are to be eliminated by
using the boundary conditions $v_x=0=v_y$ at $z=0$ and $\partial_z
v_y=0$, $\partial_z v_x + \Delta\mu p_z ({\bf r_\perp}, z=h)=0$ at
$z=h$. The resulting equations of motion for ${\bf
v_\perp}=(v_x,\,v_y)$ read
 \bea
 {\bf v_\perp}&=& \Delta\mu{c_0 \over 2\eta}
\left( {z^2 \over 2} - hz\right){\boldsymbol\nabla}_\perp
(h\partial_x^2 h) -\Delta\mu{c_0 \over 4\eta} \left( {z^3 \over 3} -
h^2z \right){\boldsymbol\nabla}_\perp \partial_x^2 h + {\kappa \over
\eta}\left( {z^2 \over 2} -hz\right){\boldsymbol\nabla}_\perp
\nabla_\perp^4 h
- \nonumber \\
&&-{\hat{C} \over \eta} \left({z^2 \over 2} -
hz\right){\boldsymbol\nabla}_\perp\partial_y \theta + {\Delta\mu
\over \eta}\left( {z^2 \over 2} -hz\right) \left[(\partial_x c +
c_0\partial_y\theta + c_0h^{-1}\partial_x h)\hat {\bf e_x} +
c_0\partial_x\theta \hat {\bf e_y}\right]. 
\label{eqxy-solid}
 \eea
Equation of motion of $h$ may be obtained by using the
incompressibility of the velocity (just as before). We obtain \bea
{\partial h \over \partial t} &=& {h_0^2 \over \eta} [{5\Delta\mu
h_0^2c_0 \over 16}q^2q_x^2h - {3c_0\Delta\mu \over 4}q_x^2h
-{4\kappa h_0 \over 3}q^6h  +
i{\hat{C}h_0 \over 3}q^2q_y\theta \nonumber \\
&&- 1/3(\Delta\mu h_0q_x^2 c + 2\Delta\mu c_0h_0q_xq_y\theta +
\Delta\mu c_0q_x^2h)] \eea The equation motion for the polar
orientation field is the same as for Model I above. After
$z$-averaging one obtains the final explicit form given by
Eq.~(\ref{thetaeq-s}) as given in Sec.~\ref{appen1}. The
concentration field obeys the same equation (\ref{conc}) as for
Model I.

\section{Dynamics and instabilities}
\subsection{Results from Model I} \label{instabilI}

Having set up the {\em effective} $2d$ coupled equations of motion
for $h({\bf q},t),\,\theta ({\bf q},t)$ and $c({\bf q},t)$, we now
examine the linear instabilities about a uniform state
$h=0,\theta=0$ and $c=0$. We begin with the extreme case with {\em
immotile but active} polar particles, i.e., $v_0=0$ but
$\Delta\mu\neq 0$. Clearly, in this limit, concentration $c$
decouples from $h$ and $\theta$.  We analyse the mode structure in
the low $\bf q$ limit. Notable characters of the mode structures are
\begin{itemize}
\item There are $O(q^0)$ {\em anisotropic} contributions proportional to $\Delta\mu
c_0/\eta$, which are of purely active origin.
 In detail: Assuming the fields to have time-dependence of the form $\exp (\Lambda t)$, we
 have for the eigenvalues $\Lambda$ of the linear stability matrix
 \bea
\Lambda(q_x,q_y) &=& {(\lambda -1)\Delta\mu c_0 q_y^2 \over 4\eta
q^2} + {(\lambda +1)\Delta\mu c_0 q_x^2 \over 4\eta q^2} -
{\Delta\mu c_0 q_x^2 \over \eta q^2}
\pm {1 \over 2}[\{ {(\lambda -1)\Delta\mu c_0q_y^2 \over 2\eta q^2} \nonumber \\
&&+ {(\lambda +1)\Delta\mu c_0 q_x^2 \over 2\eta q^2} - {2\Delta\mu
c_0 q_x^2\over \eta q^2}\}^2 - 4{(\Delta\mu)^2c_0^2 q_x^2\over
\eta^2q^4}\{ (\lambda -1)q_y^2  - {(\lambda +1)q_x^2 }\}]^{1 \over
2}. \label{eigensol}
 \eea
 Eigenvalues are anisotropic (in the Fourier space) and have
 complicated dependences on $q_x$ and $q_y$.
  It may however be noted that when
  $2q_x^2 - (\lambda-1)q_y^2/2 - (\lambda +1)q_x^2/2$ have a
  definite signature (either positive of negative) and
  $(\lambda-1)q_y^2  -(\lambda+1)q_x^2$ small, one may expand the square root in
  (\ref{eigensol}) binomially. One of the solutions
  of $\Lambda$ is then proportional to $\Delta\mu[(\lambda-1)q_y^2
  -(\lambda+1)q_x^2]/[2q_x^2 - (\lambda-1)q_y^2/2 - (\lambda +1)q_x^2/2]$ to the leading order in smallness,
  and is thus unstable just above or below the
  line $(\lambda-1)q_y^2 = (\lambda+1)q_x^2$, depending upon the
  signature of $\Delta\mu$. This is clearly indicative of instabilities
  for either sign of $\Delta\mu$. The unstable eigenvalue gets a
  particularly
  simpler form for large $|\lambda|\gg 1$. We find one of the
  eigenvalues
  \bea
 \Lambda(q_x,q_y)\sim \frac{\Delta\mu c_0q_x^2}{\eta q^4}(q_y^2 -
 q_x^2),\label{eigen2}
 \eea
 displaying clearly the instability for either signature of $\Delta\mu$
 clearly.
Further, for $q_x=0$ and $q_y=0$ one may separately obtain the
eigenfrequencies $\Lambda$ as
 \bea
 \Lambda(q_x=0,q_y)=\frac{\lambda-1}{2\eta}\Delta\mu c_0,\,0,\\
 \Lambda(q_x,q_y=0)=-\frac{2\Delta\mu c_0}{\eta} , \frac{\lambda
 +1}{2\eta} \Delta\mu c_0.
 \eea
Notice that the mathematical origin for the instability for
either signature of $\Delta\mu$ lies essentially in the mixed or
cross-coupling terms (i.e., the $h$-dependent term in the
$\theta$-equation and viceversa) and the fact that, for the leading
order in wavevector $q$, the terms on the right hand side of
Eq.~(\ref{eqh1}) have the same sign; similarly the terms on the
right hand side of Eq.~(\ref{theta}) have the same sign. Let us try
to see why this is so: There is only one source of activity in the
model, {\em viz}, the active stress given by Eq.~(\ref{stress}).
Now, this contributes to two different quantities in the {\em
effective} $2d$ descriptions: (i) $2d$ active pressure, that depends
on $h$ and (ii) $2d$ analogue of the bulk $3d$ active shear stress
that depends upon $\theta$. The dynamic of $h$ and $\theta$ depend
upon both of them. This explains the presence of the cross-coupling
terms in the dynamics. Because of the general structure of the
Stokes Eq. and our approximation of small $v_z$ (lubrication
approximation), the $2d$ active pressure and the $2d$ shear stress
come with the same signatures, the feature that is responsible for
the occurrence of the instabilities mentioned here~\cite{another}.
Since physical fields $h({\bf r})$ and $\theta ({\bf r})$ depends
upon the Fourier modes $h({\bf q})$ and $\theta ({\bf q})$ for all
$\bf q$, $h({\bf r})$ and $\theta ({\bf r})$ display instability for
both signs of $\Delta\mu$ at short enough wavenumbers. Note that
these $q$-independent contributions do not mean the modes have
finite life time at vanishing $\bf q$; such $q$-independent
behaviour is nothing but a consequence of using the Stokes' equation
(instead of the Navier-Stokes equation) for the velocity field. In
the Stokes' approximation one neglects the inertia in comparison
with the viscous term. Of course, for very low wavenumber this is no
longer valid and inertia effects will be important. Thus the above
expressions of $\Lambda$ cannot be used in the limit ${\bf
q}\rightarrow 0$. Note, however, that this does not question
the mathematical validity or consistency of the Stokesian
hydrodynamics that we have used. At the linear level, all modes
labeled by {\bf q} are decoupled. Hence, the properties of one
particular mode is unaffected by any other mode with any wavevector
(including vanishing wavevectors). Since the initial amplitudes of
the (small) perturbations considered here can be taken to zero
smoothly, the validity of the Stokesian hydrodynamics with
linearised approximation remains sound.

\item At the $O(q^2)$ the contributions
are easily presentable in the limits $q_x=0$ and $q_y=0$, when their
expressions simplify considerably. the contributions are still
purely real and display instability just like the contributions at
$q^0$. Parts of the contributions are {\em active}, parts are
however from the equilibrium part of the dynamics.  We find
 \bea
 \Lambda (q_x=0,q_y)&=& \frac{\lambda-1}{2\eta}\Delta\mu c_0 -
 {D}q_y^2,0,\label{eigenliq1}\\
 \Lambda(q_x,q_y=0)&=&-\frac{2\Delta\mu c_0}{\eta} +
 \frac{\Delta\mu c_0 h_0^2}{4\eta} q_x^2,\, \frac{\lambda
 +1}{2\eta} \Delta\mu c_0 -Dq_x^2.\label{eigenliq}
 \eea
\begin{figure}[htb]
 \includegraphics[height=7cm,angle=0]{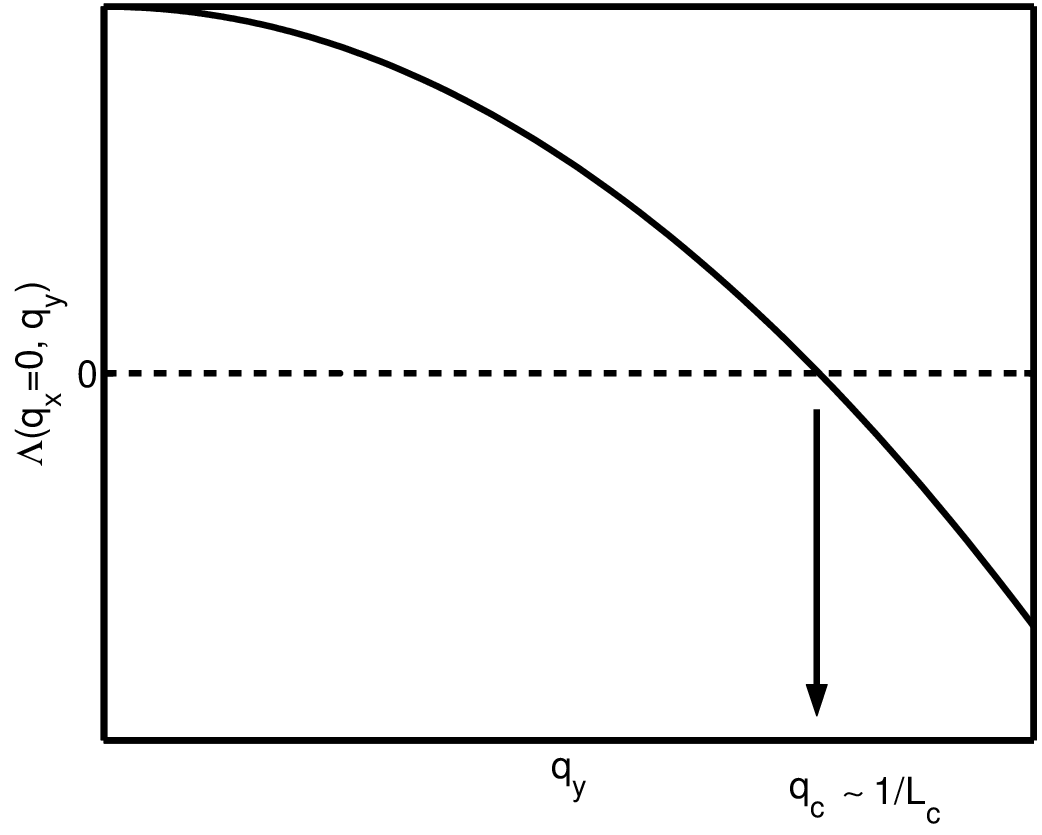} \hfill
\includegraphics[height=7cm,angle=0]{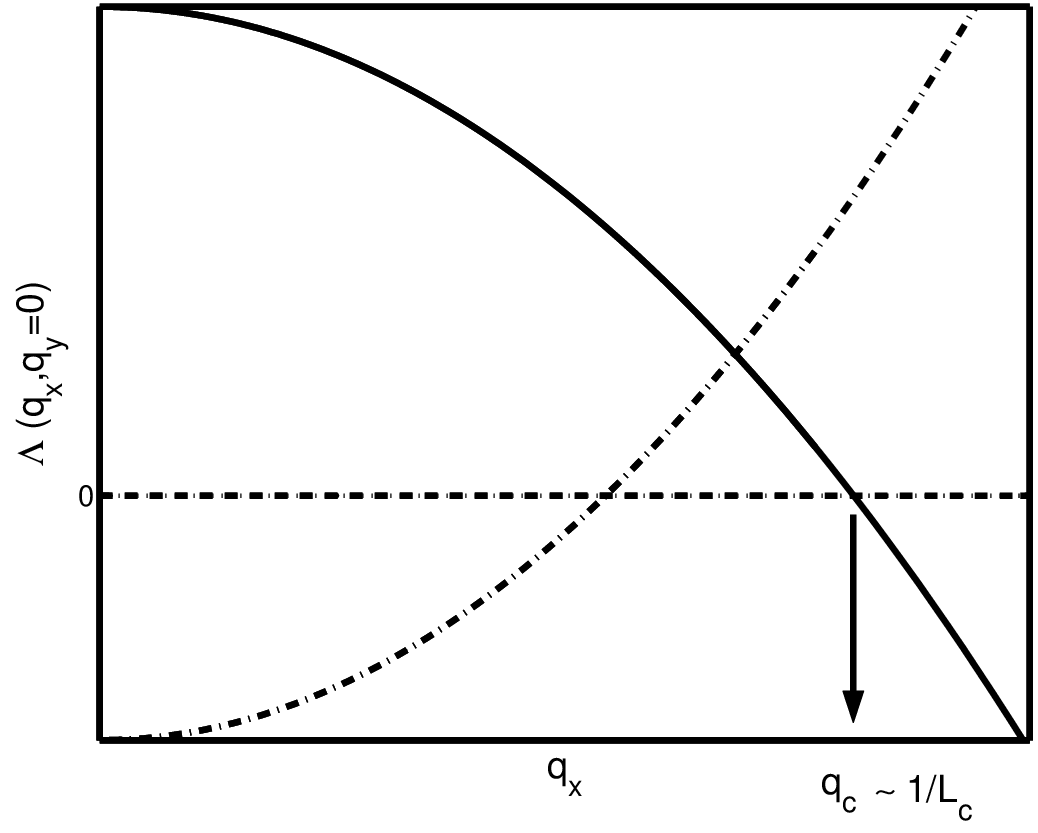}
\caption{Schematic plots showing the eigenvalues $\Lambda$ in the
two limits up to $O(q^2)$} for Model I for $\Delta\mu>0$ - {\bf
left}: $\Lambda(q_x=0,q_y)$ vs $q_y$ (a downward parabola,
continuous line) , {\bf right}: $\Lambda(q_x,q_y=0)$ vs $q_x$
(upward (dot-dash) and downward (continuous line) parabolas). The
broken horizontal line in both the plots denote $\Lambda=0$; its
meeting point with the downward parabola yields $q_c\sim 1/L_c$.
\label{eigen}
\end{figure}
A schematic diagram of the eigenvalues is given in Fig.~\ref{eigen}.
We thus see that there are both stable (coming from $D>0$) and
unstable active contributions at $O(q^2)$. Thus depending upon the
numerical value of $\Delta\mu$, there may be instabilities at
$O(q^2)$. Ignoring anisotropy, Eq. (\ref{eigenliq1}) or
(\ref{eigenliq}) allows identify a length scale
 \begin{equation}
L_c=\sqrt{\frac{D\eta}{c_0\Delta\mu}},\label{liqlc}
 \end{equation}
 such that systems with linear dimensions $L>L_c$ are unstable. Note: Due to the anisotropy,  the length scale $L_c$ should depend
 upon the polar angle which may be obtained in a more detailed calculation. Propagating modes due to the tilt-curvature coupling appear at
even higher $q$, at $O(q^{5/2})$. These propagating modes are
strongly damped at smallest wavenumbers. Finally, at a sufficiently
large $q$, all the instabilities are cut-off by the bending
stiffness $\kappa$ of the membrane.
\end{itemize}

What could be a typical value for $L_c$? We
 take linear size of an active particle $\sim$ 1nm, volume fraction
 $\phi=c_0a^3=1$ (since we assume to be in an ordered state),
 $D\eta\sim K\sim 10^{-6} dyne/cm$ where $K$ is a Frank elastic
 constant. Estimation of $\Delta\mu $ is more ambiguous:
 We use the fact that approximately $7KCal$ energy released per mole of ATP due to
 its hydrolysis. Since 1 molar mass of ATP $~\sim 500$, we obtain from its definition
 $\Delta\mu\sim 7KCal/(500gm/10^{23})$ the free energy release per unit mass per
 molecule. All these, however, lead to $L_c\sim 10 nm$, a value
 rather low compared to the linear dimensions of an eukaryotic cell.
 However, we have ignored anisotropy while estimating $L_c$.
 Moreover, for a realistic situation the effects of the ambient fluid
 is likely to be significant and should affect $L_c$.
 In any case, a direct comparison with a living cell is not very effective due to the
 simplifying approximations and the assumption of an ordered state of the active
particles that we have made.

\begin{figure}[htb]
 \includegraphics[height=7cm,angle=0]{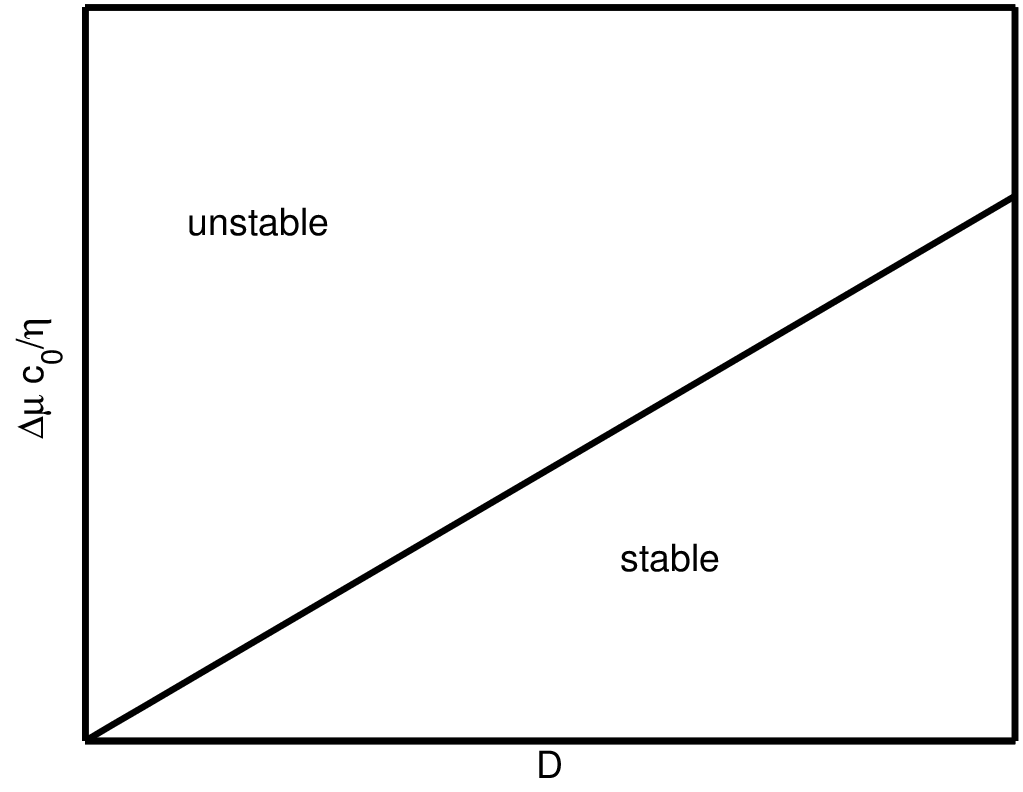} \hfill
\includegraphics[height=7cm,angle=0]{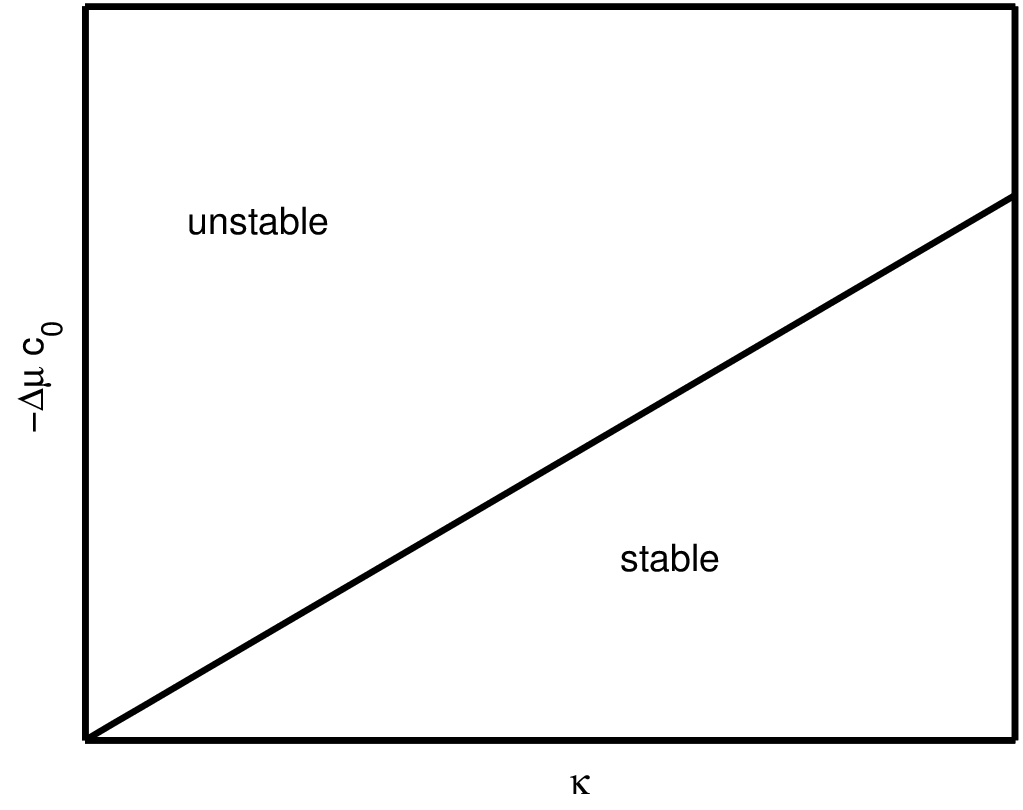}
\caption{A schematic diagram showing the unstable and stable regions
in a phase space diagram for Model I with a given $q_x$ ({\bf
left}:${\Delta\mu c_0 \over \eta}$ vs $D$ , {\bf right}: $\Delta\mu
c_0$ vs $\kappa$) for eigenvalue $\Lambda(q_x,q_y=0)$. The lines are
given by the zeros of $\Lambda (q_x,q_y=0)=0$ for a given $q_x$,
ignoring $v_0$.} \label{phase}
\end{figure}

If we now consider motile particles ($v_0\neq 0$) but continue to
ignore $c$, the most dominant effect is that one of the modes will
now pick up a propagating part with speed $v_0$ and dispersion
proportional to $q_x$. Although this is still subdominant to the
most leading order instabilities, for sufficiently large $v_0$, the
propagating modes will be observed and the instabilities at higher
$q$ corresponding to this particular mode will now be {\em moving}.
Finally, when the dynamics of $c$ is considered, the eigenmodes have
complex forms as a function of $\bf q$. A simple way to analyse them
is to look at special limits of either $q_x=0$ or $q_y=0$ (an
alternative way, considered in Ref.~\cite{sumithra}, is to examine
the limit $\Gamma \hat C/v_0\gg 1$). We find
 \bea
 \Lambda(q_x=0,q_y) &=& {(\lambda-1) \over 2\eta}\Delta\mu c_0
 -Dq_y^2\pm {1 \over 2}[ \left\{Dq_y^2 - {(\lambda-1) \over 2\eta}\Delta\mu c_0\right\}^2 \nonumber
 \\ && -4 v_0c_0\xi q_y^2]^{1/2} \;\;\; , \;\;\; -{\kappa h_0 \over
 \eta}q_y^4,\\
 \Lambda(q_x,q_y=0)&=& {2\Delta\mu c_0h_0^2 \over 4\eta}q_x^2 -
{\kappa h_0 \over \eta}q_x^4 - {2\Delta\mu c_0 \over
\eta},\;\;\;\nonumber \\ && (\lambda+1){\Delta\mu c_0 \over 2\eta} -
Dq_x^2 - ia_1v_0q_x,\;\;\;\;\;\; - iv_0q_x.\label{root2}
 \eea
Thus again in the intermediate wavenumber range there are $O(q^2)$
instabilities for sufficiently large $\Delta\mu$. In addition, there
are propagating waves coming from the concentration-orientation
coupling. Thus the instabilities and the resulting patterns will be
moving. Finally, at even larger wavenumber, all eigenmodes are
stable owing to the bending stiffness of the membrane. If
$\Delta\mu=0$, the system is stable which is expected in the
equilibrium limit. In Fig. \ref{phase} we pictorially show the
presence of instability for both signs of $\Delta\mu$ in plots of
${\Delta\mu c_0 \over \eta}$ vs $D$ and $\Delta\mu c_0$ vs $\kappa$
for eigenvalue $\Lambda(q_x,q_y=0)$ for a given value of $q_x$. The
line in the $D - \Delta\mu c_0/\eta $ plane are given by $Dq_x^2
=(\lambda+1)\Delta\mu c_0/2\eta$ for a given $q_x$ and the line in
the $\kappa - \Delta\mu c_0/\eta$ plane is given by $\kappa h_0
q_x^4 = -2 \Delta\mu c_0 + \Delta\mu c_0 h_0^2 q_x^2/2$ again for a
given $q_x$. The plots clearly indicate that for any of the
signatures of $\Delta\mu$ some regions in either of the plots
display instability.

\subsection{Results from Model II}\label{instabilII}

When there is a solid substrate, the long wavelength dynamics is
same as that in Ref.~\cite{sumithra}. We present them here for
direct comparison with the free standing case. The two eigenmodes
$\Lambda$ of the linearised coupled dynamical equations of $h$ and
$\theta$  have the form (setting $v_0=0$)
 \bea
 \Lambda &=& -\frac{11\Delta\mu c_0 h_0^2}{24\eta} q_x^2 -{\Delta\mu c_0h_0^2 \over 12\eta}q_y^2 - {Dq^2 \over 2} + {\lambda\Delta\mu c_0h_0^2 \over 6\eta}q^2 \nonumber \\
&&\pm {1 \over 2}[\{-\frac{11\Delta\mu c_0 h_0^2}{12\eta} q_x^2 -{\Delta\mu c_0h_0^2 \over 6\eta}q_y^2 - {Dq^2} + {\lambda\Delta\mu c_0h_0^2 \over 3\eta}q^2\}^2 - 8i \hat C
 \frac{\Delta\mu c_0 h_0^3}{3\eta} q_x q_y^2 - {4\Gamma \hat{C} h_0^3 \over 3\eta}q_y^2q^2 \nonumber \\
&& - {13\Delta\mu c_0Dh_0^2 \over 3\eta}q_x^2q^2 + {13(\Delta\mu)^2c_0^2h_0^4 \over 18\eta^2}q_x^4 - {13(\Delta\mu)^2c_0^2h_0^4 \over 18\eta^2}q_x^2q_y^2
+ {13(\Delta\mu)^2c_0^2h_0^4 \over 9\eta^2}q_x^2q^2]^{1/2}.
 \label{modesolid}
\eea
 As in Ref.~\cite{sumithra}, to the lowest order in wavevector,
 there is a traveling instability with a growth rate proportional
 to $q_y q_x^{1/2}$; the mode displaying
an instability travels in the $+x$ direction (-$x$ direction) if
$\Delta\mu >0$ ($\Delta\mu <0$). If we now include $v_0$, but still
ignore the dynamics of $c$, the eigenfrequencies in the limits $q_x=0$ and $q_y=0$ are
\bea
\Lambda(q_x,q_y=0)&=& -{13\Delta\mu c_0h_0^2 \over 12\eta}q_x^2 + {5\Delta\mu c_0h_0^4q_x^4 \over 16\eta} - {4\kappa h_0^3 \over 3\eta}q_x^6
\; ,\nonumber \\
&& -ia_1v_0q_x - Dq_x^2 + (2\lambda +1){\Delta\mu c_0h_0^2 \over 6\eta}q_x^2, \label{solideigen1}\\
\Lambda(q_x=0,q_y) &=& -{Dq_y^2 \over 2} + (2\lambda -1) {\Delta\mu c_0h_0^2 \over 12\eta}q_y^2 \pm
{1 \over 2}\left[ \{ Dq_y^2 - (2\lambda -1){\Delta\mu c_0h_0^2 \over 6\eta}q_y^2\}^2 -
{4\hat{C}^2 h_0^3 \over 3\eta}q_y^4\right]^{1 \over 2}. \label{solideigen}\nonumber \\
 \eea
 It is worth pointing out the basic differences between the
 eigenvalues for a free standing system given by (\ref{eigenliq1},\ref{eigenliq})
 and those for a system with a solid substrate below given by
 (\ref{modesolid}) or those in Ref.~\cite{sumithra}, whose calculational framework has been largely used here. In the former case, eigenvalues are
 $q$-independent for small $\bf q$. This is a consequence of the
 long-ranged hydrodynamic interactions. In contrast, when there is a
 solid substrate, the eigenvalues smoothly tend to zero as
 $q\rightarrow 0$. This is because any long-ranged hydrodynamic
 interactions are screened by the solid substrate below. This means, no matter how small the activity ($\Delta\mu$) is, for a
 large enough system (with linear dimension $L>L_c$) the system in Model I gets unstable. In contrast,  for model II the
 screening of the hydrodynamic interactions yields that as long as the system has a thickness smaller
 than the critical thickness $h_{0c}\sim \sqrt{\frac{D\eta}{c_0\Delta\mu}}$ for a given $\Delta\mu$, the system with arbitrarily large in-plane linear dimensions remains stable.
 This crucial
 difference apart, eigenvalues (\ref{solideigen1},\ref{solideigen}) show instabilities
 for sufficiently high $\Delta\mu$. Although the latter features are similar to that for a
 free standing system (Model I), the underlying physical mechanisms are
 different: Unlike Model I, where hydrodynamic interactions are
 responsible for type of the instabilities observed there, for Model
 II, the tilt-concentration coupling characterised by the coupling
 constant $\hat C$ \cite{sumithra} is responsible.

Regardless of the details of the boundary conditions on the system,
qualitative implications of both hold for a variety of
phenomenologies. First of all, since $v_x$ and $v_y$ depend upon
$h,\theta$ and $c$, as given in Eqs.~(\ref{eqx}) and (\ref{eqy}) for
Model I, and Eq.~(\ref{eqxy-solid}) for Model II, as soon as one of
the modes become unstable, the initial homogeneous non-flowing state
will be unstable and the active fluid starts to flow. While our
simple model calculations by themselves cannot capture the full
phenomenologies of flows in cell cortex (see, e.g.,
Refs.~\cite{salbreux,mayer}; see also Ref.~\cite{frank} for a $1d$
model for pattern formation in active fluids), our results open up
possibilities of flows in the cell cortex within calculations in a
simple setting. We show instabilities develop under generic
conditions. In addition, there may be waves under various
conditions. Our results are {\em universal} in the sense that they
do not refer to any specific cell or do not depend upon very
specific biochemical processes. Although our system and the results
which follow cannot be directly related to any biological systems
due to our several simplifying approximations, we believe our
results will inspire more realistic studies on (moving or static)
patterns ultimately formed in cell membranes (see, e.g.,
Refs.~\cite{reigada,cell-osci,dober}). Finally, in view of our
results here, let us briefly consider the possible structure of the
effective $2d$ model with a polar order that is perpendicular to the
surfaces at $z=0$ and $h$. While quantitative predictions require
detailed calculations, we can already make the following
observations, based upon the framework developed above. In this
case, again assuming a flat bottom surface (either a solid substrate
or a free active fluid surface with a very large surface tension)
the boundary conditions on the polarisation $\bf p$ are $p_z=1$ at
$z=0$ and ${\bf p}\cdot {\boldsymbol\nabla} h=1$ at $z=h$. For small
fluctuations of $p_x$ and $p_y$ (fluctuations in $p_z$ will be
second order in smallness) the boundary condition at $z=h$ implies
$p_x=\partial_x h/2,\,p_y=\partial_y h/2$. Further, $p_x$ and $p_y$
are constrained to be zero at $z=0$. Thus, in a $z$-averaged
description as above, $p_x$ and $p_y$ are slaved to
${\boldsymbol\nabla} h$. Hence, staying within the framework of $2d$
effective description as developed above, the effective dynamics
will be described by $2d$ equations of motion of height $h$ and
concentration $c$ only. This effective dynamics is now expected to
display signatures of instabilities and a spontaneous flow
transition akin to the Frederiks transition of equilibrium nematic
liquid crystals at a critical thickness for contractile active
stress with a given value of $\Delta\mu$, similar to those discussed
in Refs.~\cite{rafael,niladri1}. Details of these will be discussed
elsewhere.

\section{Diffusion coefficients}
\label{diff}

The in-plane diffusive motion of membrane bound protein are a key
ingredient in a great many biological functions including exchange
of information, material and/or energy. A complete picture of how
cells function and interact with their immediate surroundings
requires understanding of this diffusive phenomena. The recent
progress in experimental techniques to measure lateral ($D_L$) and
rotational ($D_R$) diffusion coefficients, like fluorescence
correlation spectroscopy \cite{fluro}, single particle tracking
\cite{spt}, or fluorescence recovery after photobleaching
\cite{frap}, has revealed that many of the functions performed by
proteins are crucially influenced by the diffusive behavior of the
proteins \cite{protein-diff}. Apart from its obvious biological
significance, diffusion of membrane-bound proteins is a good example
of $2d$ diffusion in a fluctuation background (here the membrane),
where the fluctuations are of nonequilibrium origin. Calculation of
$D_L$ in a strictly $2d$ flow is subtle due to the divergence
associated with the solutions of the $2d$ Stokes' equation, known as
the {\em Stokes' paradox} in the literature \cite{stokes}. In order
to overcome this Stokes' paradox, Saffman and Delbr\"uck
\cite{saffman} considered the mobility of a very thin, rigid object
in a narrow almost $2d$ perfectly flat fluid layer that is
surrounded on both sides by a further liquid and obtained finite
results for $D_L$. Calculation of $D_R$ does not suffer from any
such subtle issues. Recently in Ref.~\cite{udo} the authors
calculated $D_L$ for a protein molecule in a bare membrane. In none
of these theoretical examples, the nonequilibrium nature of the
membrane dynamics and the associated active cortical actin layer
have been considered, although the cellular cytoskeleton is often
linked with the mobility of a protein molecule in the membrane, see,
e.g., Refs.~\cite{saxton}. We here elucidate possible effects of the
cortical actin layer  nonequilibrium fluctuations on the measured
value of the diffusion coefficients within our {\em effective} $2d$
coarse-grained model. As we shall show below, the ensuing algebraic
details is rather complicated. Hence we confine ourselves here to a
treatment at the scaling level only. This suffices for our purposes
here.

Let us begin with the formal definition of the diffusion
coefficients: Since we have an anisotropic system, we consider the
general lateral diffusion tensor $D_{ij}^L$ which is defined by the
relation
 $ \langle r_i (t) r_j (t)\rangle = 2D_{ij}^L t$,
 where $r_i(t)$ is the {\em Lagrangian} coordinate of the protein molecule in the membrane at time $t$. Equating the Lagrangian
velocity of the particle with the local in-plane $2d$ velocity
field, away from the instability threshold the above definition 
leads to
$ D_{ij}^L=\frac{1}{2} \int\frac{d^2q}{(2\pi)^2}\langle v_i ({\bf
 q},\omega=0) v_j ({\bf -q},\omega=0)\rangle.$
We calculate the two lateral diffusion coefficients $D^L_{xx}$ and
$D^L_{yy}$ below. Similar to the lateral diffusion coefficients, the
rotational diffusion coefficient $D_R$ is given by
$ D_R=\frac{1}{2}\int\frac{d^2 q}{(2\pi)^2}\langle |\Omega ({\bf
 q},\omega =0)|^2\rangle,$
 where  $\Omega$ is the $z$-component of the vorticity tensor
 defined above. Thus enumeration of diffusion coefficients requires
 calculations of certain velocity correlation functions which may be done
 by stochastically driving the $2d$ equations of motion developed above. In order to
 simplify our calculations we ignore any local effect of the protein
 molecule on $\kappa$ \cite{udo}. Further, we set $v_0=0$, thus
 concentration $c$ decouples from the dynamics at the lowest order
 in $\bf q$. We present the main results below,
with some of the algebraic details available in Appendix II. For a
free standing system, we add zero-mean conserved Gaussian
distributed thermal noises in Eqs.~(\ref{eqy}-\ref{eqx}) and a
zero-mean thermal Gaussian distributed white noise in
Eq.~(\ref{thetaeq}). Finally, we calculate the diffusion
coefficients  for {\em immotile} active particles, i.e., $v_0=0$. In
addition, we ignore the concentration field $c$ in the following
calculations.

We consider Model I first. Calculations are considerably simplified
if we consider the dynamics only at the long wavelength limit, which
suffices for our purposes here. In that limit, correlators of
$h,\,\theta,\,v_x,\,v_y$ are available in Appendix II. The
noticeable feature is that  all of Eqs.~(\ref{liqcorr}) have parts
which diverge beyond a critical system size $L_c$. Ignoring
anisotropy, the length scale $L_c$ is given by Eq.~(\ref{liqlc})
above. From the correlators (\ref{liqcorr}) in Appendix II.
 we obtain for the lateral diffusion coefficients (we show only the
diverging parts below)
 \bea
 D_{xx}&\sim& \frac{K_BT}{h_0}\int \frac{d^2q}{(2\pi)^2}{ ({3\Delta\mu q_y c_0 \over \eta q^2})^2
 \over [Dq^2 + {(\lambda -1)c_0\Delta\mu \over \eta q^2}q_y^2 -
{(\lambda +1)c_0\Delta\mu \over 2\eta
q^2}q_x^2]^2}\frac{1}{\eta}[1 + \left(\frac{q_y(\lambda-1)}{2q_x}\right)],\nonumber \\
 D_{yy}&\sim& \frac{K_BT}{h_0}\int \frac{d^2q}{(2\pi)^2}{ ({\Delta\mu q_x c_0 \over \eta q^2})^2
 \over [Dq^2 + {(\lambda -1)c_0\Delta\mu \over \eta q^2}q_y^2 -
{(\lambda +1)c_0\Delta\mu \over 2\eta q^2}q_x^2]^2}\frac{1}{\eta}[1
+ \left(\frac{q_y(\lambda-1)}{2q_x}\right)]\label{liqdiff}
 \eea
Due to the complicated nature of the expressions (\ref{liqdiff}) we
do not attempt to evaluate them exactly. Instead, we treat them at
the scaling level. For a free standing system {\em without}
activity, $D_{xx}$ and $D_{yy}$ depend on the system size $L$
logarithmically, i.e., as $\log L$, which is the equilibrium
contribution. This contribution survives even when $\Delta\mu\neq 0$
which we do not show explicitly above. Evidently, for an active
system, the dependence of the active contributions to $D_{xx}$ and
$D_{yy}$ on $L$ are very different, as given by Eq.~(\ref{liqlc}).
Thus at the scaling level we obtain
 \bea
 D_{xx}, D_{yy}\sim
 \frac{K_BT}{h_0\eta}{L_c^4}(L_c^2-{L^2})^{-2}.\label{scalingD}
 \eea
  for $L\ll L_c$. Thus as $L$ increases, $D_{xx}$ and $D_{yy}$ rise
  with $L$ in power law fashion, unlike in equilibrium systems where
  such rises are logarithmic in $L$.
  Similarly,
 $D_R$, for a membrane on a thin non-active fluid film (i.e., in thermal equilibrium), has
 no dependence on $L$, it instead depends on the small-scale cut-off
 $l$ (of the order of the particle size) as $l^{-2}$. In contrast,
 when the fluid film is active, $D_R\sim \frac{K_BT}{\eta
 h_0}\frac{L_c^2}{(L_c^2-L^2)^2}$ below the threshold. Thus, similar
 to $D_{xx}^L$ and $D_{yy}^L$, $D_R^L$ increases as $L$ increases.
 For a thin inflexible isotropic fluid layer with viscosity $\eta$ surrounded
 by a bulk fluid of viscosity $\eta_1$ with $\eta_1/\eta$ finite,
 theoretical calculations of Ref.~\cite{saffman}
 diffusion coefficient of an inclusion of a small but finite size
 is finite. Our work is thus an extension of Ref.~\cite{saffman},
 incorporating effects of orientation degrees of freedom and a fluid
 membrane, but staying at the special simple limit of
 $\eta_1/\eta\rightarrow 0$.
 In addition, our assumption of inflexible active fluid-isotropic bulk fluid interface
 will not strictly hold in a real cell. Thus more refined
 calculations are needed for better quantative estimations. We conclude this Section by mentioning that
 there are two important physical effects which we have not considered. First of all, it is well-known that
 for a membrane with a finite thickness in thermal equilibrium,
 fluctuations increase the {\em effective thickness}
 \cite{nir1} which in turn reduces the diffusivity.
 Similar effects should be observed in the present problem as well,
 whose quantitative enumeration requires further work which we do
 not discuss here. However, our results above holds
 in the limit of
 zero membrane thickness (the parameter $h_0$ here corresponds to
 the average thickness of the active fluid layer, and not
 of the thickness of the membrane at the top). Secondly, in view of the findings of Ref.~\cite{naji} that
 a quenched rough surface reduces the effective diffusion coefficient substantially, whereas, for
 an annealed surface the reduction is relatively small, it would be important to investigate the analogous
 effects in the present problem. However, direct application of the results of Ref.~\cite{naji}
 to our problem is difficult due to the presence of additional degrees of freedom
 (polarisation fluctuation $\theta $) with long-ranged correlations and the active stress. Further work is
 necessary to settle this issue properly. This is beyond the scope of the present work.


In order to calculate diffusion coefficients  of an inclusion for
Model II we need to set $\hat c=0$, since otherwise there are
underdamped propagating modes in the system. From the correlators
(\ref{corr-sol}) we find that, unlike the free standing system,
there are no critical lateral size beyond which instabilities set
in; further their dependence on $L$ is $\log L$. Instead, now there
is a critical thickness $h_{0c}$ such that for a film with thickness
larger than $h_{0c}$, $h_{0c}\sim \sqrt{\frac{D\eta}{c_0\Delta\mu}}$
spontaneous flow instabilities akin to the Frederiks transition in
equilibrium nematics set in \cite{rafael,niladri1}. For a system
with $h_0 <h_{0c}$ we find
 \bea
 D_{xx}^s,\,D_{yy}^s&\sim&{K_BT \over \eta h_0}\frac{h_{0c}^4}{(h_{0c}^2-h_0^2)^2}\log L
 \eea
 for $h_0$ smaller than the critical thickness $h_{0c}$. Thus $D_{xx}^s$ and $D_{yy}^s$
 depend explicitly on $h_0$, a measure of the system size. Note the
 differences between the expressions of diffusion coefficients in
 Model I and II. For Model I, they depend upon the linear size $L$
 of the system in a way markedly different from the $\log L$
 dependence as observed in equilibrium systems. In contrast, in
 Model  II, they depend on $L$ as $\log L$; in addition however they
 acquire non-trivial dependences on $h_0$.

\section{Summary and outlook}
\label{conclu}

In this article we have set up $2d$ coarse-grained equations for a
coupled system of a fluid membrane and a thin layer of active
cortical actins anchored to it in terms of a height field,
orientation field and concentration of active particles. We
considered two cases of a free standing system and a solid substrate
under the system, separately. The distinguishing feature of the
former case is the presence of long ranged hydrodynamic
interactions. We discuss the generic instabilities and patterns
which appear due to activity. We use our equations to calculate
lateral and rotational diffusion coefficients of an inclusion in the
membrane. For the case of a free standing film, we have assumed that
the surface tension of the bottom free surface is large, so that
fluctuations of that surface are suppressed and are not considered
in the subsequent calculations. This is mainly a theoretically
interesting case and for real experimentally testable systems,
surface tension of the surfaces should be finite and hence the
bottom surface will have fluctuations. Despite the limitations of
our approximations, our results bring out the differences between
the two cases (free standing system and system resting on a solid
surface) very clearly. For the sake of analytical convenience we
have ignored a few details of cell membranes, e.g., effects of the
surface tension of the membrane, the role of active proteins in the
membrane and local modulation of the bending stiffness due to the
inclusion; these may however be important in a real biological set
up. Qualitative features of our basic results should in principle be
testable in standard cell biology experiments measuring, e.g.,
measurements of diffusion coefficients. However, direct comparisons
with {\em in vivo} experiments will not be easy primarily due to our
assumptions of idealised model and also due to many complicated
features of a cell membrane (e.g., a cell membrane is actually a
bilayer or the cortical actin is anchored to the membrane only at
discrete junctions). Our work here relates to Ref.~\cite{saffman} in
that we include new effects coming due to orientational degrees of
freedom, active stress and finite membrane stiffness, but consider
the limiting case where the ambient fluid viscosity is much smaller
than the active fluid viscosity. Despite the limitations of our
simplifying assumptions, out work shows the dramatic effects of
hydrodynamic interactions and provides a first step towards more
realistic calculations. We hope our work will stimulate more
realistic calculations in this direction.

Apart from its phenomenological importance, we believe our work is a
first step in formulating and understanding the $2d$ coupled
dynamics of a fluid membrane and driven orientational broken
symmetry modes. It would be interesting to see how the predictions
of Ref.~\cite{toner} get modified due to the fluctuations of the
membrane. Secondly the fluctuating orientational degrees of freedom
 should create an effective long ranged interactions (of nonequilibrium origin) between
 different parts of the membrane. It would be interesting to see
 whether such interactions may allow a finite temperature crumpling
 transition of a $2d$ fluid membrane \cite{niladri3}, something which is prohibited in
 equilibrium \cite{weinberg}.

\section{Appendix I: Full equations for $\theta$}
\label{appen1}

In order to obtain an {\em effective} two-dimensional description of
the dynamics of $\theta$ for a free standing system (Model I), we
average over the $z$ direction to get
  \bea
{\partial \theta \over \partial t} &=& -ia_1v_o q_x \theta  - i\xi
q_yc - Dq^2\theta -{\lambda \over \eta q^2}
\left[ {3\Delta\mu c_0h_0 \over 4}q_yq_x^3h + i\hat{C}q_xq^4\theta - \Gamma\kappa q_xq_yq^4h\right] - i\hat{C}q_yq^2h \nonumber \\
&&+ {(\lambda-1) \over 2\eta q^2h_0}[\Delta\mu h_0q_xq_yc +
h_0c_0\Delta\mu q_y^2\theta + 2c_0\Delta\mu q_xq_yh] +
 {(\lambda+1)c_0\Delta\mu \over 2\eta q^2}q_x^2\theta.
\label{thetaeq}
 \eea
The corresponding dynamical equation for $\theta$ when there is a
solid substrate below  (Model II) is \bea
\partial_t\theta &=& -ia_1v_0q_x\theta - i\xi q_y c + i\Gamma \hat{C}q_y h - Dq^2\theta + {7(\lambda-1) \over 24\eta}c_0\Delta\mu h_0 q_xq_yh - {\lambda \over 16\eta}\Delta\mu c_0h_0^3 q_yq_x^3h \nonumber \\
&&+ {\lambda\kappa \over \eta}h_0^2q_xq_yq^4h - i{\hat{C} \over 3\eta}h_0^2q_y^2q_x\theta + {(\lambda-1) \over 6\eta}h_0^2\Delta\mu q_xq_yc + {\lambda \over 3\eta}\Delta\mu c_0h_0^2 q^2\theta - {\Delta\mu \over 6\eta}c_0 h_0^2q_y^2\theta \nonumber \\
&&+ {\Delta\mu \over 6\eta}c_0h_0^2q_x^2 \theta + {\lambda\sigma
h_0^2 \over 3\eta}q^2q_xq_y h. \label{thetaeq-s}
 \eea

\section{Appendix II: Correlation functions}
\label{appenII}
\subsection{Model I: Free standing system}
Equations (\ref{eqx}) and (\ref{eqy}) reduce to
 \bea
 v_x&=& -\frac{i}{\eta q^2 h_0}[c_0h_0\Delta\mu q_y\theta +
 2c_0\Delta \mu q_x h],\\
 v_y&=& -\frac{i}{\eta q^2} q_x\theta. \label{shortI}
 \eea
The  equations for $\theta$ and $h$ in the Fourier space in the long
wavelength limit are
  \bea
 i\omega\theta + Dq^2\theta -\frac{\lambda -1}{2\eta q^2 h_0}(h_0c_0 \Delta\mu
 q_y^2 \theta + 2 c_0\Delta\mu q_x q_y h) - \frac{\lambda +1}{2\eta
 q^2} c_0\Delta\mu q_x^2\theta &=& g_\theta, \\ \label{shortItheta}
 i\omega h - \frac{1}{\eta q^2}(2c_0 h_0 \Delta\mu q_x q_y \theta +
 2\Delta\mu c_0 q_x^2 h) &=& g_h,\label{shortIh}
 \eea
 where $g_\theta$ are $g_h$ are zero-mean Gaussian distributed white
and conserved noises respectively. In the above we have ignored the
dynamics of $c$ and set $v_0=0$. Equations (\ref{shortItheta}) and
(\ref{shortIh}) may be solved in a straight forward way to obtain
the corelators $\langle |h({\bf q},\omega)|^2\rangle$ and $\langle
|\theta ({\bf q},\omega)|^2\rangle$. The resulting expressions are
lengthy and not very illuminating; we do not present the full
expressions here. Instead, we obtain the correlators in the long
wavelength limit and use them to obtain correlations functions
$\langle |v_x({\bf q},\omega)|^2\rangle,\,\langle |v_y({\bf
q},\omega)|^2\rangle$ and $\langle|\Omega ({\bf
q},\omega)|^2\rangle$ (where $\Omega ({\bf q})$ is the Fourier
transform of the $z$-component of ${\boldsymbol\nabla}\times {\bf
v}$). We find(ignoring parts which show no divergence)
 \bea
\langle|h({\bf q},\omega=0)|^2\rangle &=& 2K_BT/\eta{({4h_0 q_y \over 3q_x})^2
[1+\{{(\lambda -1)q_y \over q_x}\}^2] \over [Dq^2 + {(\lambda -1)c_0\Delta\mu \over 2\eta
q^2}q_y^2 - {(\lambda +1)c_0\Delta\mu \over 2\eta
q^2}q_x^2]^2}  \nonumber \\
\langle|\theta({\bf q},\omega=0)|^2\rangle &=&  2K_BT/\eta{[1 +
\{{(\lambda -1)q_y \over q_x}\}^2] \over [Dq^2 + {(\lambda
-1)c_0\Delta\mu \over 2\eta q^2}q_y^2 - {(\lambda +1)c_0\Delta\mu
\over 2\eta
q^2}q_x^2]^2}, \nonumber \\
\langle |v_x({\bf q},\omega=0)|^2\rangle &=&({3\Delta\mu q_y c_0
\over \eta q^2})^2\frac{2K_BT}{\eta}{1 + \{{(\lambda -1)q_y \over
q_x}\}^2 \over [Dq^2 + {(\lambda -1)c_0\Delta\mu \over 2\eta
q^2}q_y^2 - {(\lambda +1)c_0\Delta\mu \over 2\eta
q^2}q_x^2]^2},\nonumber \\
\langle |v_y({\bf q},\omega=0)|^2\rangle &=&({c_0\Delta\mu q_x \over
\eta q^2})^2\frac{2K_BT}{\eta}{1+  \{{(\lambda -1)q_y \over
q_x}\}^2 \over [Dq^2 + {(\lambda -1)c_0\Delta\mu \over 2\eta
q^2}q_y^2 - {(\lambda +1)c_0\Delta\mu \over 2\eta
q^2}q_x^2]^2}, \nonumber \\
\langle|\Omega({\bf q},\omega=0)|^2\rangle &=& \left( {c_0\Delta\mu
(q_x^2-3q_y^2) \over \eta q^2}\right)^2\frac{2K_BT}{\eta}{1+
\{{(\lambda -1)q_y \over q_x}\}^2 \over [Dq^2  + {(\lambda
-1)c_0\Delta\mu \over 2\eta
q^2}q_y^2- {(\lambda +1)c_0\Delta\mu \over 2\eta q^2}q_x^2]^2}.
\label{liqcorr}
 \eea
 Evidently, all the
correlators diverge beyond a typical critical system size $L_c$
defined above.

\subsection{System in contact with a solid substrate}

The stochastically driven $2d$ equations of motion of
$h,\,\theta,\,v_x$ and $v_y$ are
 \bea
 v_x&=& - i\frac{c_0\Delta\mu h_0}{\eta}q_x h -
 i\frac{\Delta\mu c_0 h_0^2}{2\eta}q_y\theta + \xi_x^s, \nonumber \\
 v_y&=&-i\frac{\Delta\mu c_0 h_0^2}{2\eta}q_x\theta +\xi_y^s, \nonumber \\
 \frac{\partial h}{\partial t}&=& -{13c_0\Delta\mu h_0^2 q_x^2 \over 12\eta}h - {2h_0^3c_0\Delta\mu q_xq_y \over 3\eta}\theta +\xi_h^s,\nonumber \\
 \frac{\partial\theta}{\partial t}&=&-Dq^2\theta
 + {7(\lambda -1) c_0h_0\Delta\mu q_xq_y \over 24\eta}h + {\lambda c_0h_0^2 \Delta\mu q^2 \over 3\eta}\theta - {\Delta\mu c_0h_0^2 q_y^2 \over 6\eta}\theta + {\Delta\mu c_0h_0^2 q_x^2 \over 6\eta}\theta
 +\xi_\theta^s. \nonumber \\
 \label{shorteqsol}
 \eea
 Here the noises $\xi_x^s,\,\xi_y^s,\,\xi_h^s$ and $\xi_\theta^s$
 are all zero-mean Gaussian white noises.
 Equations (\ref{shorteqsol}) lead to the velocity correlators
 \bea
 \langle |v_x ({\bf q},\omega=0)|^2\rangle &=& ({9\Delta\mu c_0h_0^2 q_y \over 78 \eta})^2{\langle \xi_\theta^s({\bf q},\omega=0)\xi_\theta^s(-{\bf q},\omega=0)\rangle + \{{7(\lambda -1)q_y \over 26h_0q_x}\}^2\langle |\xi_h^s({\bf q},\omega=0)|^2\rangle \over \left[Dq^2 - {\lambda\Delta\mu c_0h_0^2 \over 3\eta}q^2 - {\Delta\mu c_0h_0^2 \over 6\eta}q_x^2 + {c_0\Delta\mu h_0^2 \over \eta}\{{7(\lambda -1) \over 39} + {1 \over 6}\}q_y^2\right]^2} \nonumber \\
&&+ \langle \xi_x^s({\bf q},\omega=0)\xi_x^s(-{\bf q},\omega=0)\rangle, \nonumber  \\
 \langle |v_y ({\bf q},\omega=0)|^2\rangle &=& ({\Delta\mu c_0h_0^2 q_x \over 2\eta})^2{\langle \xi_\theta^s({\bf q},\omega=0)\xi_\theta^s(-{\bf q},\omega=0)\rangle + \{{7(\lambda -1)q_y \over 26h_0q_x}\}^2\langle |\xi_h^s({\bf q},\omega=0)|^2\rangle \over \left[Dq^2 - {\lambda\Delta\mu c_0h_0^2 \over 3\eta}q^2 - {\Delta\mu c_0h_0^2 \over 6\eta}q_x^2 + {c_0\Delta\mu h_0^2 \over \eta}\{{7(\lambda -1) \over 39} + {1 \over 6}\}q_y^2\right]^2} \nonumber \\
&&+ \langle \xi_y^s({\bf q},\omega)\xi_y^s(-{\bf q},\omega=0)\rangle
 \label{corr-sol}
 \eea
Thus there are no instabilities at any finite lateral dimension of
the system, and hence there is no critical $L$, unlike the free
standing system.
\section{Acknowledgement}
The authors thank J.-F. Joanny and S. Ramaswamy for fruitful
discussions and suggestions. AB gratefully acknowledges partial
financial support in the form of the Max-Planck Partner Group at the
Saha Institute of Nuclear Physics, Calcutta funded jointly by the
Max-Planck- Gesellschaft (Germany) and the Department of Science and
Technology (India) through the Partner Group programme (2009).


\begin{thebibliography}{99}
\bibitem{foot1} Hence for  wavevector range given by $\eta q^2 \sim \eta' q/h_0$, i.e., when the ambient
fluid is expected to become important, our results are not expected
to hold.
\bibitem{foot-vari} Thermal noise $\xi_\theta$ has variance
proportional to the inverse of the {\em rotational viscosity}, which
we assume to be numerically same as $\eta$ for simplicity.
\end{thebibliography}
\end{document}